\documentclass[seceq]{ptptex}

\usepackage{graphicx}
\usepackage{wrapft}

\notypesetlogo                       

\title{ Universal Short-Range Repulsion in the Baryon System Originating from the Confinement}        

\subtitle{Approach in String-Junction Model   
 \footnote{Preliminarily reported in Ref. \citen{SokenarXiv}.}
}

\author{
Ryozo {\sc Tamagaki}\footnote{E-mail: tama-ktn@nike.eonet.ne.jp} 
}

\inst{
Kamitakano Maeda-Cho 26-5, Kyoto 606-0097, Japan \\
}

\abst{
Aim of this study is to show a way to unifiedly understand the origin of repulsive core of baryon-baryon interaction and 
the origin of universal repulsion of three-baryon interaction whose importance has recently become noticeable. 
``Universal" means its effect independent of flavor and spin of baryons. 
As shown in recent lattice QCD calculations, the confinement mechanism in the single baryon is realized in the Y-shaped structure 
of the strings embodying the squeezed color flux-tubes that consists of a junction at the central position and 
three oriented strings linking the quarks with the junction. Here we adopt a string-junction model suitable 
for description of such baryonic structure, in order to find out substantial  features that the confinement mechanism persists 
in multi-baryon systems when baryons closely approach to each other. First we clarify an originating mechanism of the repulsive core 
of baryon-baryon interactin, and next by extending this approach to the three-baryon system, we derive a universal short-range 
three-body repulsion. Key concept in the study lies in the recognition that such short-range repulsion appears as latent effect implying 
the energy necessary for full overlapp of baryons, that is, the energy for creation of the junction and anti-junction pair which 
leads to specific exotic baryon states of gluonic excitation.        
The short-range interaction derived in such a way is universal, because the effect comes from the color degrees of freedom. 
This three-body repulsion prevents the equation of state from the dramatic softening due to hyperon-mixing 
in neutron-star matter, and thus we can disolve the difficulty that otherwise theory contradicts with the observations on 
the masses of neutron stars.

}

\begin{document}

\maketitle

\newcommand{\gsim}{>\kern-12pt\lower5pt\hbox{$\displaystyle\sim$}}
\newcommand{\lsim}{<\kern-12pt\lower5pt\hbox{$\displaystyle\sim$}}
\newcommand{\mbf}{\boldsymbol}
\newcommand{\cd}{$\cdot$}  

\section{Introduction}

In recent studies of neutron star (NS) matter with a hyperon-mixed core at high density, Takatsuka et al. \cite{NYT02} have shown 
that universal repulsion of three-baryon interaction (3BI) is indispensable to avoid the dramatic softening of equation of state 
(EOS) due to the hyperon-mixing, which leads to the contradiction to the observations on the masses of NSs.\footnote{We use 
the following abbreviations; NS for neutron star, 3BI for three-baryon interaction, EOS for equation of state, 
and SJM for string-junction model.}  
Here ``universal" means that the repulsion is independent of flavor and spin. This recognition is significant, 
although phenomenological, because the statement holds almost independently of the models describing NS matter. 
Importance of the recognition is similar to the case of the repulsive core model for nuclear force 
proposed by Jastrow in 1951,\cite{Jastrow} who succeeded in explaining the high energy nucleon-nucleon scattering data at that time 
without violation of the charge-independence of nuclear force. Therefore it is a very interesting problem to study 
the origin of such a universal 3BI repulsion. In three-body force, there is 3BI of meson-exchange nature 
acting in the intermediate region as well as 3BI acting at small inter-baryon distances. In this paper we concentrate on  
3BI of short-range nature.  

Recently we have studied the problem whether or not the 3BI due to the two-pion exchange via $\Delta$-excitation gives rise a 
universally repulsive effect in high-density neutron star matter, and obtained the following results.\cite{TNTRCNP}  
Although this 3BI acts repulsively in pure neutron matter, it does not provide a repulsive effect at high density 
strong enough to prevent the EOS of NSs from the dramatinc softening due to the hyperon-mixing, because it does not acts 
on the admixed $\Lambda$-particle due to its spin-isospin dependence. Even though there may be other possibilities of 
mesonic origin, to get universally replusive effects in 3BI is quite open. 

Noticing the property that the repulsive 3BI needed phenomenologically is common for all the baryons, 
we study the problem from the viewpoint of the quark confinement, the nonpertubative effect inherent in the color degrees of freedom in QCD.
In treating the baryon system, we should adopt a model which well describes the confinement mechanism in the single baryon. 
In recent studies with use of the lattice QCD calculations, Suganuma et al.\cite{TSNM02} have shown that the confinement mechanism  
in the single baryon is represented as the Y-shaped structure consisting of a junction at the central point and 
three oriented strings embodying the squeezed color flux-tubes. The junction plays a role to neutralize the color and 
the strings stem from the junction and end at the quarks.  Such a Y-shaped string-junction picture of the single baryon 
was firstly pointed out by Nambu in 1973,\cite{Nambu73} and adopted as one of key structural elements in the string-juntion model 
developed in the 1970's \cite{PTPS78}. The recent lattice QCD calculations have given the basis that this picture for the confinement 
in the single baryon is real. In the present paper we study the problem by adopting the string-junction model we abbreviate as SJM.

At first, developing a previous study based on the SJM by the author in 1982\cite{Tamagaki82}, we show a way to understand the originating 
mechanism of the repulsive core in the two-baryon system, where we assume the existence of exotic dibaryon states predicted by the SJM. 
Extending such a line of approach to three baryon system, we derive the universal repulsion of 3BI with short-range nature in the SJM. 
In the next step we study the effects from this part of 3BI on the EOS of neutron matter, by using the framework constructing an    
effective density-dependent two-body potential from 3BI. Because the resulting 3BI is universal, its effect in the NS matter does not 
change with the hyperon-mixing.  Finally we show that the EOS obtained by including the effects from this universal repulsion of 3BI  
is similar to the sufficiently stiff EOS given by Takatsuka et al.\cite{TNYT06}. Thus we can avoid the drasmatic softening of EOS 
due to the hyperon-mixing which otherwise leads to the contradiction with the observations on NS masses. 

This paper is organized as follows. After a brief description on the string-junction model in \S 2, we show the way 
to understand the origin of the $BB$ repulsive core in \S 3. In \S 4, the derivation of the universal repulsion of 3BI in the SJM 
is described, the density-dependent effective two-body potential is derived from this 3BI repulsion and results of numerical 
calculations are shown. Then in \S 5, the effects on the EOS of NS matter are numerically calculated and the results are shown. 
Finally several related points are discussed in \S 6. Concluding remarks are given in the last section.

\section{Brief description of the string-junction model (SJM)}
In this section, we briefly explain the SJM and give mass formula of hadrons which are used in later sections. 

\subsection{Description of ordinary hadrons in the SJM}
The SJM was developed in the later half of the 1970's, following the studies of hadron reactions with use of quark lines and 
their rearrangement diagrams.\cite{Okubo78,PTPS78} A series of studies on the SJM by Imachi, Otsuki and Toyoda
\cite{IOT757677,IO77,IO78} are to be especially noted. Study of the present paper is performed on the basis of these works. 
In the SJM the color flux-tubes embodying the confinement mechanism are illustrated by strings with orientation, as in Fig.~1.  
For mesons, we define the orientation of strings as stemming from the anti-quark ($\bar{q}$) and ending at the quark ($q$). 
(The reverse choice is equally possible, as this is matter of definition.) The confinement mechanism is represented 
by formation of a closed net of oriented string(s). An ordinary meson is drawn by a single string as shown for M in Fig.~1. 
An ordinary baryon (anti-baryon) is drwan by the Y-shaped net with three strings stemming 
from a junction $J$ (ending at an anti-junction $\bar{J}$), as shown for $B$ ($\bar {B}$) in Fig.~1. The arrow attached to the string
represents its orientation that means the predisposition of the order in appearance of $q$ (denoted by $\circ$) and $\bar{q}$ 
(denoted by $\bullet$) when we cut the string. 
This holds also for the case that $q$ and/or $\bar{q}$ do not appear at the end. Following our definition, 
$q$ appears ahead of the arrow and $\bar{q}$ appears behind the arrow. 
We simply write the ordinary meson of $(q\bar{q})$ as $M$, the ordinary baryon of $(qqqJ)$ as $B$ and the ordinary anti-baryon 
$(\bar{q}\bar{q}\bar{q}\bar{J})$ as $\bar{B}$, where $(\cdots)$ denotes the color singlet object. 

\begin{figure}[htbp]
\begin{center}
\includegraphics[width=0.9\linewidth]{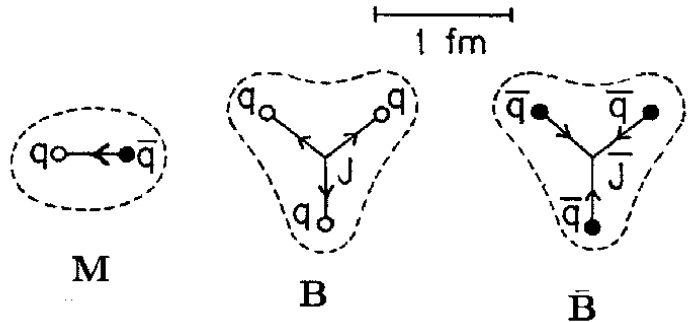}
\end{center}
\caption{Pictorial view of the ordinary hadrons; $M$, $B$ and $\bar{B}$. The figures do not represent individual states of hadrons, 
but rather imply the states from which actual states are generated as Regge trajectories after the angular momentum projection, 
like the so-called intrinsic states in deformed nuclei. When we superpose various rotated configulations in the body-fixed frame, 
there appear the states spreading smoothly. The dotted curves illustrate the region supposed in the bag model.}
\label{fig 1}
\end{figure}

The string structure for $M$ has been noticed as a suitable represention for the color flux-tube since the early stage of QCD study.  
For the structure of $B$, Suganuma and others have shown, in their recent lattice QCD calculations, that the junction structure of 
the Y-shaped strings is real.\cite{TSNM02,TMNS01} This recognition has given the sound basis to such a view adopted in the SJM 
studies in the 1970's. The results of these QCD simulations are summarized in the energy expression of a potential form\cite{TSNM02}:

\begin{eqnarray}
V_{q\bar{q}}(r)=\sigma_{q\bar{q}}\;r-\frac{A_{q\bar{q}}}{r}+C_{q\bar{q}}, \\
V_{3q}=\sigma_{3q}L_{\rm min}-A_{3q}\sum_{i<j}\frac{1}{|{\mbf r}_{i}-{\mbf r}_{j}|}+C_{3q}, \\
\sigma_{3q}\simeq 0.89\;{\rm GeV/fm}\simeq \sigma_{q\bar{q}},\;\;A_{3q}\simeq\frac{1}{2}A_{q\bar{q}}.
\end{eqnarray}
Here $r$ for $M$ is the distance between $q$ and $\bar{q}$, and $L_{\rm min}$ for $B$ is the minimum of the total lentgh 
of the color flux-tubes linking three quarks with $J$.  The 1st term of $V_{3q}$ is the energy of the strings 
with the string tension almost equal to that of the $q$-$\bar{q}$ case; this means the universality of the string tension. 
The 2nd term represents the color Coulomb-type attraction of the one-gluon exchange nature, as inferred from the relation 
$A_{3q}\simeq\frac{1}{2}A_{q\bar{q}}$. Such presentation has been shown to be valid over the actual range of 
$L_{\rm min}\sim (0.5-1.5)$ fm. The lattice QCD calculations have shown the profile of this Y-shaped string-junction in the action 
density,\cite{Ichie03c} and the actual length of the strings from the junction to the quarks, $\ell_s\simeq 0.5$ fm,
\cite{TSIMNO03c} that is, $L_{\rm min}\simeq 1.5$ fm for the equilateral configuration of the quarks.          
 
\subsection{Mass formula for hadrons in the SJM}
Usually the hadrons other than $M$, $B$ and $\bar{B}$ are called exotic hadrons. This terminology is somewhat ambiguous, 
because there possibly exist many kinds of hadronic states, which are not simply described in the form of 
$(q\bar{q})$, $(qqqJ)$ and $(\bar{q}\bar{q}\bar{q}\bar{J})$ but not necessarily exotic, e.g., molecular states of the ordinary hadrons. 
In the SJM, we can descriminate exotic hadrons from nonexotic hadrons; exotic hadrons  are characterized by the appearance of 
the interjunction string (abbreviated as $IJ$) which connects $J$ and $\bar{J}$. 
Hadrons are constructed by three kinds of building blocks, that is, the quark $q$ (anti-quark $\bar{q}$) with one oriented string, 
the junction $J$ (anti-junction $\bar{J}$) with three strings and the string of the inter-junction $IJ$. 
We define the total number of quarks (the sum of 
the numbers of $q$ and $\bar{q}$) denoted as $N_q$, the total number of junctions (the sum of the numbers of $J$ and 
$\bar{J}$) by $N_J$ and the total number of interjunctions by $N_{IJ}$. Then we denote the energy of each building block by
 $m_q$, $m_J$ and $m_{IJ}$, respectively, where we take the same mass for $q$ and $\bar{q}$ and the same energy for $J$ and $\bar{J}$.  

The mass formula in the SJM are given in the simplest form, as follows:\cite{IO78}
\begin{equation}
m(N_q,N_J,N_{IJ})=m_q N_q+m_J N_J+m_{IJ}N_{IJ}=m_B N_J-\delta N_{IJ},
\end{equation}
where we use the relations,
\begin{equation}
N_q=3N_J-2N_{IJ}\;\;(N_J\neq 0),\;\;m_B\equiv 3m_q+m_J,\;\;\delta\equiv 2m_q-m_{IJ}. 
\end{equation}
The mass $m_q$ represents the sum of the current quark mass and the energy of an attached string. For the light quarks 
($u$, $d$ and $s$) the string energy is the main part of $m_q$. In this paper we treat such a case. The mass $m_B$ represents the average 
of the ground state of the single baryon. If $|\delta|$ is much smaller than the baryon mass $m_B\sim 1$ GeV, the energies of hadrons are mainly 
given by $m_B N_J$ and spread by the secondary contribution $-\delta\:N_{IJ}$.  Estimate of magnitude of $\delta$ is discussed later. 
 For the time being, we proceed under the assumption of the property $|\delta|\ll m_B$, which is proved in \S 6.

The most importnat point in this mass formula lies in the property that the order of a hadron mass is determined by the number of 
the junctions as $m(N_q,\;N_J,\; N_{IJ})\sim N_J$ GeV, which we preferentially use in the following arguments.

\begin{figure}[htbp]
\begin{center}
\includegraphics[width=0.9\linewidth]{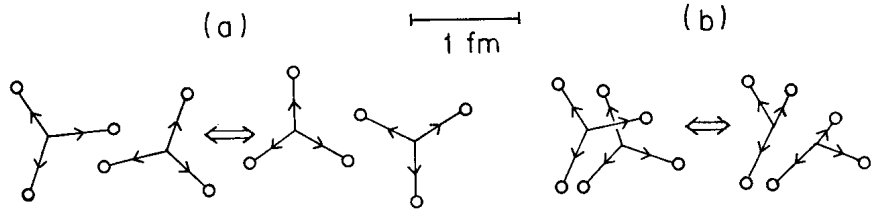}
\end{center}
\caption{ Possible features of strings in approaching two baryons; (a) two adjacent strings    
change their connections with the nearest junctions to make the total length of strings minimum (flip-flop prpcess) 
and (b) at closer approach, two $B$ suffer strong deformation but the strings still keep the Y-shaped structure.　　　}
\label{fig 2}
\end{figure}

\subsection{Excitation of ordinary hadrons}
A typical order of magnitude for increase of energy of $M$ and $B$ is considered as that of the masses of vector mesons such as 
$\rho$ and $\omega$. Meson production as fission of the string takes place for the elongation of a string with the order of 
$0.5-0.7$ fm in $M$ and $B$. Excitation with such order of magnitude is also possible by deformation of the flux-tube and 
by swing in a relative angle between strings, which are regarded as vibrational modes in a broad sense. These are the excitations 
keeping the basic structure of $M$ and $B$, and the substatial features of the confinement do not change. 

At close approach of two baryons, when two Y-shaped strings begin to overlap, the rearrangement between two adjacnet strings  
takes place to make deviation of the total length of the strings minimum, as shown in Fig.~2(a), which is called 
the ``flip-flop" process of strings.\cite{Miyazawa79} At closer approach, strong deformation of the Y-shaped strings necessarily 
occurs so as to keep the Y-structure, as illustralted in Fig.~2(b). Resulting excitation energy is supposed to be considerably large. 
            
\section{Origin of repulsive core of baryon-baryon interaction described in the String-Junction model}

At present it is an open problem whether the repulsive core\footnote{There are several ways to describe the short-range repulsive 
effect in the $BB$ interaction. We simply call this effect by ``repusive core" in this paper.} exists commonly in all the $BB$ states or 
changes its aspects depending on flavor and spin.  At least it certainly exists in the ${}^1 S_0$  and ${}^3 S_1$ states 
in the nucleon-nucleon ($NN$) system. Although there is no sound evidence for its existence in the $NN$ $P$-waves and higher partial waves, 
the assumption of its existence in all the $NN$ states does not contradict with the experimental data. 
As for the systems involving hyperons ($Y$), there is no experimental information on specific properties of short-range $NY$ and $YY$  
interactions. Most of theoretical studies are usually done following the $NN$ case under the assumption of approximate falvor symmetry.        
At present the assumption that the repulsive core exists in all the states of the $BB$ system does not contradict with the observations. 
\footnote{If the so-called H-particle, a dibaryon state of the flavor singlet, would be observed as a spatially compact bound state, 
this assumption does not hold. But at present the H-particle has not been observed. }

 From the viewpoint of the SJM, because we consider the repulsive core as an effect originating from the confinement mechanism, 
the universal repulsive core (acting commonly for all the baryon pairs) is a natural consequence.

\subsection{Exotic dibaryon states and junction-pair states}
When two $B$ approach very closely, the two Y-shaped strings inevitablly overlap. At full overlap 
two $B$ lose their identity because they cannot keep the Y-shaped structure and finally fuse.
How does the confinement mechanism work at such situation? To this problem, the SJM provides us with the viewpoint that 
the strings and junctions make the closed net of the color flux-tubes with six quarks at the ends by producing the exotic 
dibaryon state, which is illustrated in the (left) of Fig.~3. As it has $N_q=6$ and $N_J=4$, we denote it by $D_6^{4}$.
\footnote{We denote a hadron $H$ with $N_q$ and $N_J$ by $H_{N_q}^{N_J}$.}

In this exotic dibaryon state $D_6^{4}$, there appear three $J$ and one $\bar{J}$, and the three interjunctions connecting them, 
which are absent in the ordinary hadrons, manifest the exotic character. Because the reality of $J$ and consequently 
 of $\bar{J}$ becomes firm, the appearance of the oriented interjunction connecting them (indicated by the arrow emerging from $J$ and 
ending $\bar{J}$) is naturally expected. Thus we consider the existence of $D_6^{4}$ quite reasonable. 
Although the auguments in the following done on the assumption of the existence of $D_6^{4}$ predicted by the SJM are speculative, 
we can obtain meaningful consequence. In the last two sections, after presentation of the results, we make comments on 
such problematic points that positive observational information on the existence of $D_6^{4}$ has not been reported yet and 
no lattice QCD simulation about it has been performed yet.
\newpage
The mass of $D_6^{4}$ is expressed with use of Eqs.~(2\cd 4) and (2\cd 5):
\begin{eqnarray}
m(D_6^{4})
&=& 6m_q+3m_J+m_{\bar{J}}+3m_{IJ}=2(3m_q+m_J)+m_J+m_{\bar{J}}+3m_{IJ} \nonumber \\
         &=& 2m_B+{\cal E},\\
{\cal E}&\equiv& 2m_J+3m_{IJ}.
\end{eqnarray}

\begin{figure}[t]
\begin{center}
\includegraphics[width=0.9\linewidth]{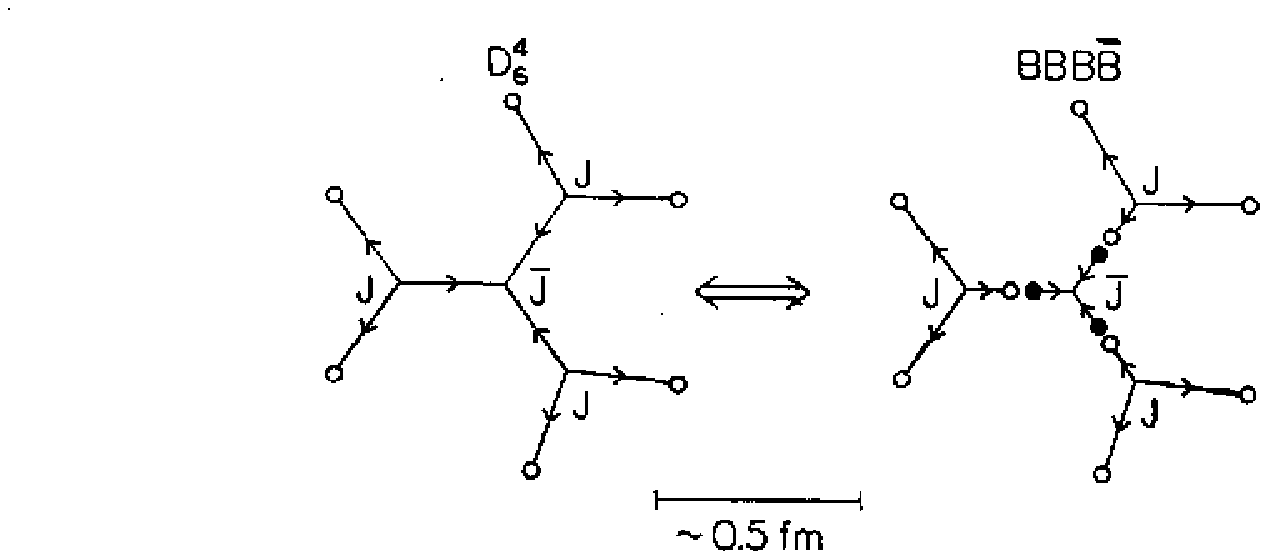}
\end{center}
\caption{ (Left）String-junction structure of exotic dibaryon $D_6^{4}$. (Right） Formation of 
$BBB\bar{B}$ through fission of three interjunctions.}
\label{fig 3}
\end{figure}

The energy of $D_6^{4}$ is higher than that of the ordinary state of $2B$ ($2m_B$) by ${\cal E}$. 
$D_6^{4}$  can not turn back directly to $BB$. In order to turn into a system of the ordinary hadrons, the cut of three interjunctions is required. 
In doing so, as shown in Fig.~3 (right), at the cut places the quark pairs $q\bar{q}$ are created and then the baryon-pair $B\bar{B}$ 
is created. As a result the system turns to $BBB\bar{B}$. Its energy is higher than $2m_B$ by about $2m_B$, and therefore 
we can take ${\cal E}\sim 2$ GeV. \footnote{In the isolated $B\bar{B}$, meson-exchange effect works attractively 
mainly due to the $\omega$ and $\sigma$, but here such effect is not considered because it is uncertain in 
many-baryon configuration under consideration.}
This is simply understood from the fact that $N_J$ increases by 2, compared with the ordinary 2$B$, according to the mass formula.

\begin{figure}[t]
\begin{center}
\includegraphics[width=0.9\linewidth]{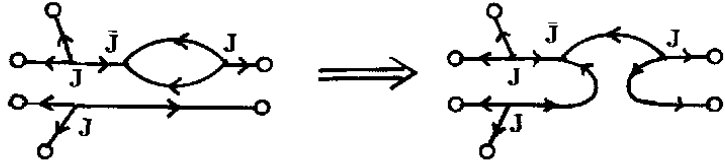}
\end{center}
\caption{ Illustration of an example for another process to create $D_6^{4}$ ; (left) an excited string appears to creat    
a string loop and a $J\bar{J}$ pair and another adjacent string is in parallel with the same orientation which is opposite 
to the orientation of the strings in the loop and (right) rearrangement takes palce between two strings with opposite directions, 
leading to the formation of $D_6^{4}$.}
\label{fig 4}
\end{figure}

In the description mentioned above, the exotic dibaryon $D_6^{4}$ is considered as to be formed through the virtual creation of 
the $B\bar{B}$ pair.  The exotic dibaryon $D_6^{4}$ is also formed through another process. Such an example is shown in Fig.~4.  
In the (left) of this figure, situation proir formation is illustrated: In order for two parallel strings with the same orientation 
belonging to different baryons to overlap, one of them is excited by formation of the string loop accompanying the junction pair. 
In the (right) of the figure, situation just after formation of $D_6^{4}$ is illustrated: Two strings with opposite orientations 
transform into two hair-pin like strings.\cite{Nambu74} The $D_6^{4}$ is surely formed, although some of the strings are 
strongly deformed. In this case too, since the junction number $N_J$ increases by 2 for the formation of one junction pair, 
the excitaion energy is about 2 GeV. Such a loop-type excitation of one string has been studied in a recent work on 
the pentaquark problem by Suganuma and others.\cite{SOTI05,SIOT04} In lattice QCD calculations in a study of  
the single baryon, gluonic excitation has been shown to appear at about 1 GeV and to be regarded as a global excitation of 
the whole Y-type flux-tube system.\cite{TS04} On the contrary, the excitation shown in Fig.~4 is that of a 
single flux-tube and appears in the energy near ${\cal E}\sim 2\;$GeV for the $B\bar{B}$-type excitation, 
apart from the deformation energy of the string.

The $B\bar{B}$ pair turns into a series of states commonly involving the $J\bar{J}$ pair, as shown in Fig.~5; 
$M_4^{2}$ and $M_2^{2}$（exotic mesons), and states involving no quark $S_0^{2}$（a kind of glue balls).
We note that, in later descriptions, ${\cal E}\sim 2$GeV plays a role of the key number. The loop in $M_2^{2}$ created by fusion of the 
$q$ and $\bar{q}$ in $M_4^{2}$ is the same with the one appearing in Fig.~4. So we may call the excitations mentioned above simply 
the $B\bar{B}$ excitation or the $J\bar {J}$ excitation. 

This series of states with the baryon number=0 and $N_J=2$ contain commonly the $J$ and $\bar{J}$ which we call the  
junction-pair. Here they are denoted by the symbol ${\cal M}$. Because of irrelevance to the number of quarks, 
they are the states inherent in the gluon degrees of freedom. The energies of ${\cal M}$ are distributed about the central value of 
about 2 GeV with a spread of several hundreds MeV, as far as we are concerned with $u$, $d$ and $s$. (For details, \S 6.1 should 
be referred to.) This means that there exist a group of states with the baryon number=2, consisting of $D_6^{4}$ and 
[$BB$+ ${\cal M}$], in the energy region higher by about 2 GeV above the ordinary $BB$.

\begin{figure}[htbp]
\begin{center}
\includegraphics[width=0.9\linewidth]{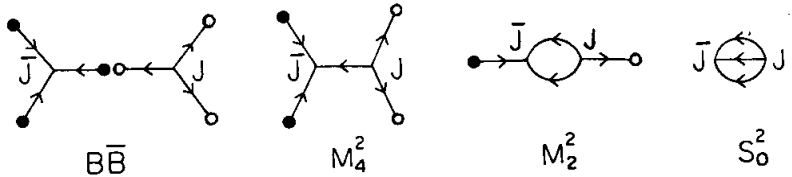}
\end{center}
\caption{ A series of states with the baryon number =0 formed by fusion of  $q$ and $\bar{q}$ from $B\bar{B}$, which 
we call junction-pair states, denoted as ${\cal M}$.}
\label{fig 5}
\end{figure}

\subsection{Repulsive core of the $BB$ interaction}
We can paraphrase the statement mentioned above as follows. In order for the confinement mechanism to persist when two baryons 
overlap fully, it is necessary to form the string-junction net where the arrows attached to the oriented strings are closed.
It requests the excitation energy of ${\cal E}\sim 2$ GeV, which makes the creation of $B\bar{B}$ pair (more generally the junction pair) 
possible. If the kinetic energy of $BB$ in the center of mass system is much smaller than $m_B$ (e.g., in the elastic scattering region), 
the system is apt to avoid such a large excitation.  The potentiality that two baryons reside apart to get rid of such expense 
is equivalent to the existense of a high potential barrier between two baryons, namely, the repulsive core. 
Its height is of the order of ${\cal E}\sim 2$GeV, and its range is the relative distance where two baryons begin to overlap largely. 

 The potential mentioned above may be defined in a sense of the adiabatic approximation usually adopted to get a static 
potential, where the relative kinetic energy of two baryons is taken into account in solving the Shr\"odinger equation. 
In principle, we should treat the energy described in the quark-gluon level of QCD, but here we use the approximate value
 obtained in the SJM, abbreviated as $E^{(SJM)}$. We define an interaction  potential dominating in the core region to be used in 
the baryon level, by the difference between the energies $E^{(SJM)}$ at two relative distances of $BB$, $r_{12}=|{\mbf r}_1-{\mbf r}_2|$ and 
$r_{12}\rightarrow\infty$, where ${\mbf r}_1$ and ${\mbf r}_2$ are the positions of two baryons: 
\begin{subequations}
\label{eq:1}
\begin{equation}
V^{({\rm core})}(r_{12}) \equiv E^{(SJM)}(r_{12})-E^{(SJM)}(r_{12}\rightarrow \infty) .
\label{eq:1a}
\end{equation}
For the SJM, since we can give the values of $E^{(SJM)}$ only at the two limiting cases, at $r_{12}\ll \ell_s$ and $r_{12}\gg \ell_s$, 
compared with the string length of the ordinary baryon $\ell_s\simeq 0.5$ fm,
the magnitude of $V^{({\rm core})}$ at closest approach is obtained as  
\begin{eqnarray}
V^{({\rm core})}(r_{12}\ll \ell_s) \simeq E^{(SJM)}(r_{12}\ll \ell_s)-E^{(SJM)}(r_{12}\gg \ell_s)\\
\simeq E(D_6^{4})-2m_B\simeq 2m_B+{\cal E}-2m_B\simeq {\cal E}.
\end{eqnarray}
\end{subequations}
This $V^{({\rm core})}$ is regarded as the interaction of short range in the hadron level, and is incorpolated into the two-baryon potential, 
together with the meson-exchange potentials of the long and intermediate range.

The view that the core-like short-range repulsion is a latent effect reflecting the energy needed for the overlapp is 
quite natural because it is familar in the atomic and nuclear systems. 
The repulsive core between two atoms (typically between two closed-shell atoms), which is often approximated 
by a hard-sphere potential, is understood as originating from the large excitation energy of electron states which is needed for 
the overlap of the electron clouds. This repulsive core is ``strong"; it is so steep that its core height 
becomes by several order of magnitude larger than the strength of attractive pocket just ouside the core, if a soft core 
potential is used. The repulsive core used in the effective potential between two nuclei 
(typically between two light closed-shell nuclei, e.g., between two $\alpha$-particles) is understood as originating from 
the excitation energy of nucleon states required for overlap of two nuclei. This repulsive core is ``moderately strong",  
as its height of $\sim $ 150 MeV is larger by one order of magnitude than the depth of attractive pocket ($\sim $ 10 MeV)  
for the $\alpha-\alpha$ case, although a hard core potential may be used phenomenologically for low energy scattering. 
\footnote{This repulsive soft core in the potential form is understood as a reflection of the forbidden relative states 
which should vanish under the Pauli-exclusion principle. Such a feature is described by using the orthogonality condition to the 
forbidden relative states or by using a strong nonlocal potential.} 
The $BB$ repulsive core of the $\sim $2 GeV height with the $\sim 0.1$ GeV attractive pocket (in the $S$-wave) belongs to 
the  ``moderately strong" class. 

These features of the repusive core in the atomic and nuclear systems are attibuted to the effects of the Pauli-exclusion principle. 
The related internal degrees of freedom is the spin of the electron in the atomic system and the spin-isospin 
of the nucleon in the nuclear system. As for the repulsive core of the $BB$ system, the related internal degrees of freedom is 
the color. Although we see similarity among three cases, it is to be noted that
the confinement mechanism is essential in the $BB$ case, and the exclusion principle relevant to 
the color degrees of freedom alone is not essential because the quarks have many internal degrees of freedom.

The range of the repusive core originating from the confinement mentioned above is smaller than the relative distance  
where two Y-shaped structures begin to touch. If the string length of the Y-shape in the ground state of $B$ is taken as 
$\ell_s\simeq 0.5$ fm correspoding to $L_{\min}\simeq 1.5$ fm in Eq.~(2\cd2),  
the half overlap of two $B$ corresponds to the relative distance $\simeq$0.5 fm. Thus the range of about 0.5 fm is a reasonable 
choice for the repulsive core derived in the SJM. In a previous work in 1964,\cite{OTW64} where we considered  
``structural core" of the $NN$ interaction from the viewpoint of the composite model of hadrons, we have chosen the shape of the $NN$ 
repulsive core as the gaussian form, implying that the repulsion grows as the overlap of two nucleons becomes large.  
On the same standpoint, the present author constructed a realistic $NN$ potential with the gaussian repulsive core, 
called OPEG.\cite{OPEG} In a representative version of OPEG potential, OPEG-A, the strength and the range of the repulsive core 
are taken as 
\begin{equation}
{\cal V}_C(r)={\cal V}_C^{0} \;{\rm exp}\left[-(r/{\eta}_C)^{2}\right],\;\;{\rm with}\;{\cal V}_C^{0}=2\;{\rm GeV},
\;\;{\eta}_C= (0.45-0.5)\; {\rm fm}.
\end{equation}
The repulsive core $V^{({\rm core})}$ obtained in the SJM has the feature well corresponding to ${\cal V}_C$ of OPEG-A. Thus we can understand 
the origin of the $BB$ repulsive core in the SJM, as the manifestation of the confinement mechanism in the $BB$ system. 

In the notion of the state vector, the present context is interpretated as follows.  The state vector of the system $|\Psi\rangle$ is 
a superposition of the part spanned by the states of two ordinary baryons, denoted as $P|\Psi\rangle$, and the part spanned 
by the $D_6^{4}$ plus $\left[ BB+{\cal M}\right ]$ states, denoted as $Q|\Psi\rangle$, where $P$ and $Q$ are the projection operators ($P+Q=1$). 
The latter states $Q|\Psi\rangle$, whose existence is limited to the core region and at high excitation, has the latent influence 
on the former state $P|\Psi\rangle$.  Such influence appears as suppression of the $BB$ relative wavefunction at very small distances, 
corresponding to the effects by the repulsive core. 

It is not simple to show logical connection between the core-like repulsion shown in the adabatic approximation and the effective 
$BB$ interaction described in the $P|\Psi\rangle$ influenced by the existence of $Q|\Psi\rangle$.  
First of all we notice that substantial effects appear in two ways. One is the effect of the orthogonality between $P|\Psi\rangle$ 
being in the scattering state and $Q|\Psi\rangle$ with spatially compact sructure and high excitation energy. The other is 
the channel coupling effect. The resulting core-like repulsion in the adiabatic approximation means that the former overwhelms the latter. 
The reasoning is as follows.
\begin{enumerate} 
\item {\it Effect of the orthogonality} \\ 
In the scattering problem, as the energy is fixed to the 
incident energy, the observable effect appears in scattering phase shifts in the asymptotic region. In the problem under consideration, 
the influence from $Q|\Psi\rangle$ is rather special, because its existence is limited in the inner-most region of 
the $BB$ relative distance (core region), $r\lsim \;b\simeq 0.5$ fm. We notice that the primary influence from 
the orthogonality to $Q|\Psi\rangle$ brings about the nodal behavior of the $BB$ relative 
radial wavefunction of $P|\Psi\rangle$ at the core region, which is to be described by a strong nonlocal interaction. 
The reason is that such radial nodes do not mean attraction to produce bound states but the almost energy-independent 
nodal behavior gives repulsive effects, because the amplitude of the oscillatory wavefunction is suppressed and 
the outermost node plays a role equivalent to the boundary condition given by a core-like repulsion with the radius at $r\simeq b$, 
in off-resonance situation.\footnote{Previously we showed such an example in the $NN$ ${}^1 S_0$ scattering.\cite{OTY65,Tamagaki67}}
If there exist many such states of $Q|\Psi\rangle$, the wavefunctions are strongly suppressed 
in the whole core region and the situation becomes as if the hard core exists. If some attractive potential works outside 
the core region, even though the scattering phase shifts show attractive feature at low energies, repulsive feature appears 
at high energies because repulsive-core effects are strong enough to overwhelm the attractive effects. 
The repulsive core is to be considered as a simple local potential form in such a limited sense that it represents only 
the suppressed feature of the wavefunction given by the nonlocal interaction in the core region mentioned above. 
\footnote{We have encountered such problem concerning interaction between nuclei:
The strong nonlocal interaction giving rise to the inside oscillatory wavefunction appears when we treat the inter-nucleus coordinate 
as a dynamical variable with use of the resonating group method, while the repulsive core with literal meaning appears when we 
treat the inter-nucleus distance as a variatianal parameter with use of the generator coordinate method.\cite{HT72,Saito77} 
} 
\item {\it Channel coupling effect} \\
In addition there appears a dynamical effect due to the channel coupling through the interaction matrix elements (i.m.e.) 
between $P|\Psi\rangle$ and $Q|\Psi\rangle$. Generally the channel coupling effect is considered to give attractive effect on 
the channel with a lower energy, in ordinary situation of bound states. Since the states of $P|\Psi\rangle$ and $Q|\Psi\rangle$ 
under consideration have substantially different structures with the large energy difference, the i.m.e. are not large, and 
 the channel coupling effect seems moderate. We cannot enter further into this problem, because of lack of concrete knowledge 
about the wavefunctions of $Q|\Psi\rangle$ in the SJM.
\end{enumerate}

We can summarize the results in the SJM approach to the $BB$ short-range interaction, by saying that the repusive core preventing 
two baryons from close approach appears as the latent effect on the two-baryon states, corresponding to the high excitation 
energy needed for the system to fully overlap.

\section{Short-range three-body repulsion in the $BBB$ system in the string-junction model}

In this section, with use of the SJM, at first we seek main aspects of three-baryon interaction (3BI) originating from the confinement,  
next give its potential form and strength and then derive a density-dependent effective two-body potential which is used 
to obtain its contributions to the energy of baryonic matter. Before entering individual steps, we briefly describe 
the treatment of 3BI adopted here.

\subsection{Outline of treatment of three-baryon interaction in baryonic matter}

The Hamiltonian, which enables us to understand properties of nuclei and baryonic matter, is expressed in the nonrelativistic treatment 
as follows:
\begin{equation}
H=\sum_i K_i+\sum_{i<j}V(i,j)+\sum_{i<j<k}V^{(3B)}(i,j,k)+\cdots,
\end{equation}
where $K_i$ is the kinetic energy of the $i$-th baryon $B_i$ (including the rest mass difference from the nucleon mass),  
$V$ two-baryon interaction and $V^{(3B)}$ three-baryon interaction. The ellipsis means four-body (and higher) interactions, 
which are neglected. We take the following form for $V^{(3B)}(i,j,k)$ consisting of the cyclic sum:   
\begin{equation} 
V^{(3B)}(i,j,k)=\sum_{{\rm cyclic}}W(i,j;k)=W(i,j;k)+W(j,k;i)+W(k,i;j),
\end{equation}
where $W(i,j;k)$ represents the 3BI which arises additionally from the approach of the $k$-th baryon to the pair of 
$B_i$ and $B_j$. Such a form is familiar, for example, in the study of the three-nucleon interaction from 
the two-pion exchange process via $\Delta$-excitaion of the 3rd nucleon $N_3$ intervening between two nucleons 
($N_1$ and $N_2$), initiated by Fujita and Miyazawa.\cite{FujitaMiyazawa} 
(We take $i=1,\;j=2,\;k=3$ as the representative one without loss of generality.)  
Also the repulsive three-nucleon interaction introduced phenomenologically by Phandharipande and others has such a form.\cite{AP97,HParnps00} 
As shown below, this form is suitable for the 3BI derived in the SJM.\footnote{The cyclic sum is naturally understood by viewing the situation 
shown in Fig.~6 of \S 4.2. The 3BI under consideration in the SJM is regarded as arising through the virtual creation of one $B\bar{B}$ pair 
in $B_3$ approaching to the two baryons $B_1B_2$ near by. The formal structure in this case is analogous to the case 
of Fujita-Miyazawa type three-nucleon interaction, if the $\Delta$-excitaion in the latter is supposed to be replaced by the one 
$B\bar{B}$ pair excitaion in the former. The expression of Eq.~(4\cd2) implies a wider situation and is applicable to other cases 
not restricted to the cases mentioned above.}  
A primary problem is how to describe a representative one of the SJM type, $W(1,2;3)$. 

Once $W(1,2;3)$ is obtained in the SJM, next we derive the effective two-body potential by taking the sum of the diagonal matrix elements 
over the the occupied states of the 3rd baryon $B_3$ in the baryonic matter. This effective two-body potential importantly depends on the baryon 
density $\rho$, and we denote it by $U(1,2;\rho)$. At this step we take into account the effects of the short-range correlation 
(abbreviated as SRC) coming from the interactions between $B_1-B_3$ and between $B_2-B_3$, while the SRC between $B_1-B_2$ is 
included in the final step of the energy calculation. Then we can calculate the energy of the baryonic matter $E$ in the framework of 
the reaction matrix theory, by adopting $V(i,j)+U(i,j;\rho)$ as the two-body potential. In this way the symmetric sum of $V^{(3B)}(i,j,k)$ 
in Eq.~(4\cd2) is automaticlly included in the sum over all the two-baryon pairs.  The interaction energy $E_{\rm int}$ is given by 
\begin{equation}
E_{\rm int}=\frac{1}{2}\sum_{\alpha,\beta}^{\rm occupied}\langle \phi_{\alpha}(1)\phi_{\beta}(2)|V(1,2)+U(1,2;\rho)
|\psi_{\alpha,\beta}(1,2)-\psi_{\alpha,\beta}(2,1)\rangle,
\end{equation}
where  $\phi_{\alpha}(1)\phi_{\beta}(2)$ is the unperturbed (plain-wave) function of the baryon pair and $\psi_{\alpha,\beta}(1,2)$ is 
the pair wave function in the matter which includes two-body correlations. In addition to the contribution from the usual two-body potential $V$,  
the effects of the 3BI in the SJM is given as the contribution from the the effective two-body potential $U$. 
\begin{figure}[b]
\begin{center}
\includegraphics[width=0.8\linewidth]{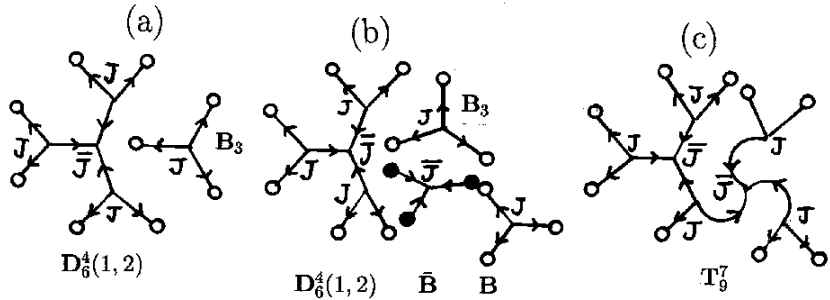}
\end{center}
\caption{Illustration of formation of the exotic tribaryon in two steps; (a）approach of $B_3$ to the exotic dibaryon $D_6^{4}(1,2)$ 
formed by adjacent $B_1$ and $B_2$, (b) virtual creation of one $B\bar{B}$ pair between them, and (c) formation of the exotic 
tribaryon $T_9^{7}$ after the fusion of the strings of $q$ and $\bar{q}$.}　　　
\label{fig 6}
\end{figure}

\subsection{Exotic tribaryon states in the $BBB$ system and strength of three-baryon force}
Here in the SJM, we study the representative part of the 3BI, $W(1,2;3)$, which arises additionally  
when the 3rd baryon $B_3$ approaches to the adjacent two baryons ($B_1$ and $B_2$). We show that a short-range three-body repulsion 
originates from the confinement mechanism, by extending the reasoning in deriving the universal repulsive core of the $BB$ interaction 
in the previous section to the three-baryon system.    

In order for baryons to overlap fully, from the viewpoint of the SJM, the strings and junctions representing color flux-tubes 
should be rearranged to hold the confinement in such situation. In the $BB$ system, it becomes possible by creating 
one $B\bar{B}$ pair. Here we consider the extension of this idea to the BBB system by taking two steps.  
(1) We imagine the situation that $B_1$ and $B_2$ are close enough to form the exotic dibaryon state $D_6^{4}(1,2)$ and the third baryon 
$B_3$ approaches to $D_6^{4}(1,2)$, as illustrated in Fig.~6(a). (2) One $B\bar{B}$ pair is virtually created between them as in Fig.~6(b), 
and after the fusion of the strings of $q$ and $\bar{q}$ the system turns to the connected string-junction net with $N_q=9$ and $N_J=7$ 
with increase of three interjunctions instead of three $q\bar{q}$-pair strings, as illustrated in Fig.~6(c). This state is 
the exotic tribaryon state denoted by $T_9^{7}$.  
 
\begin{figure}[b]
\begin{center}
\includegraphics[width=0.8\linewidth]{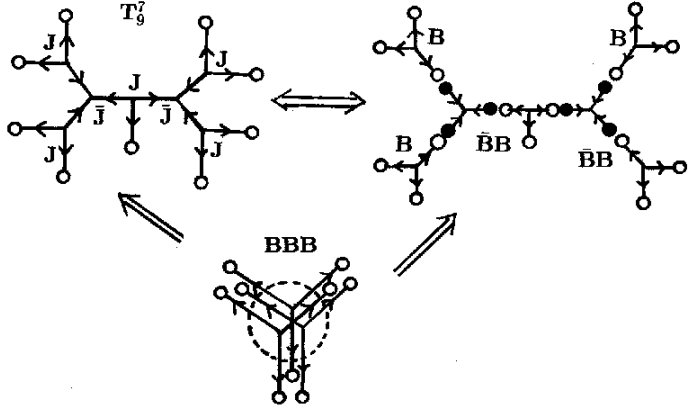}
\end{center}
\caption{Pictorial view of the exotic tribaryon and its preformation stage; (left）string-junction net of the tribaryon $T_9^{7}$,  
 (right）$BBB(\bar{B}B)(\bar{B}B)$ states arising from the fission of the interjunction strings, and 
(bottom) an example illustrating one of many possible configurations for full overlap of $BBB$, 
where the dotted area indicates the formation region of a string-junction net. Such view is the unfolded-sheet drawing of the tribaryon 
having three-dimensional spread.}　　　
\label{fig 7}
\end{figure}

The energy of $T_9^{7}$ is given using the mass formula as follows:
\begin{eqnarray}
m(T_9^{7})&=& 9m_q+5m_J+2m_{\bar{J}}+6m_{IJ}=3(3m_q+m_J)+2m_J+2m_{\bar{J}}+6m_{IJ} \nonumber \\
&=&3m_B+4m_J+6m_{IJ}=3m_B+2{\cal E}.
\end{eqnarray}
The mass $m(T_9^{7})$ is higher by $2{\cal E}\sim 4$ GeV than the energy of the ordinary $BBB$ system, where ${\cal E}$ is 
defined in Eq.~(3\cd2). In this energy, its half (${\cal E}\sim 2$ GeV) is attributed to the formation of $D_6^{4}(1,2)$. 
Therefore the energy arising additionally from the approach of $B_3$ is ${\cal E}\sim 2$ GeV. This is clear, because 
the $B\bar{B}$ is created and the junction number increases by 2.  Next we consider the 3BI based on the SJM, whose strength is 
this additional increase of energy (${\cal E}\sim 2$ GeV) at the closest approach of $B_3$ to $D_6^{4}(1,2)$. 
  
We write the 3BI in the SJM (abbreviated as SJM-3BI), following the reasoning adopted in the two-baryon case. 
A representative part of the SJM-3BI denoted by $W(1,2;3)$ is given as the difference between the energies at the two situations 
taken along the ``path" which leads to the formation of $T_9^{7}$ at the close approach of $B_3$ to the exotic dibaryon $D_6^{4}(1,2)$. 
These energies are to be described in the quark-gluon level of QCD, in principle, but here we use the approximate values obtained 
in the SJM, denoted by $E^{(SJM)}(\xi_3;D_6^{4}(1,2))$, where $\xi_3$ is the relative distance between the position of $B_3$ 
and that of the center of mass of $D_6^{4}$. We define the magnitude of $W(1,2;3)$ as 
\begin{subequations}
\label{eq:1}
\begin{equation}
W(1,2;3)\equiv  E^{(SJM)}(\xi_3;D_6^{4}(1,2))-E^{(SJM)}(\xi_3\rightarrow \infty;D_6^{4}(1,2)),
\label{eq:1a}
\end{equation}
For the SJM, since we can give the values of $E^{(SJM)}$ only at the two limiting situations, namely, those at $\xi_{3}\ll \ell_s$ 
and $\xi_3\gg \ell_s$ (equivalent to $\xi_3 \rightarrow \infty$), compared with the string length of the ordinary baryon 
$\ell_s\simeq 0.5$ fm, the magnitude of $W(1,2;3)$ at the closest approach of $B_3$ is obtained as  
\begin{eqnarray}
 & W(1,2;3)& \simeq E^{(SJM)}(\xi_3\ll \ell_s;D_6^{4}(1,2))-E^{(SJM)}(\xi_3\gg \ell_s;D_6^{4}(1,2)) \\
 &\simeq&E(T_9^{7})-(m_B+E(D_6^{4}))\simeq 3m_B+2{\cal E}-(m_B+2m_B+{\cal E})\simeq{\cal E},  
\end{eqnarray}
\end{subequations}
which is considered as the strength of the SJM-3BI. 

$W(1,2;3)$ regarded as the interaction in the hadron level is a genuine 3BI in the following sense.
In $D_6^{4}(1,2)$, the original baryons ($B_1$ and $B_2$) have lost their identity of the baryon as shown in Fig.~6(a), 
although $D_6^{4}(1,2)$ is formed from $B_1$ and $B_2$ through the virtual creation of the $B\bar{B}$ pair. Therefore, 
the interaction between $D_6^{4}(1,2)$ and $B_3$ in the configuration leading to 
the formation of $T_9^{7}$ is an intrinsic three-body effect which is not reduced to any effect due to the two-body interaction. 

The state vector of the system $|\Psi\rangle$ can be written, with use of two projection operators $P$ and $Q$ satisfying $P+Q=1$, 
as the sum of the state vector $P|\Psi\rangle$ which describes the states of three ordinary baryons and the state vector 
$Q|\Psi\rangle$, which describes the  states arising from $D_6^{4}(i,j)$ plus $B_k$ and leading to $T_9^{7}$ 
(denoted by $Q_T|\Psi\rangle$) and the higher states with larger junction numbers. Restricting $Q|\Psi\rangle$ to $Q_T|\Psi\rangle$, 
we can write $Q_T$ as $Q_T=Q_3+Q_1+Q_2$, where $Q_3$ projects out the states brought about along the ``path" leading to $T_9^{7}$ formed 
from $D_6^{4}(1,2)$ and the intervener $B_3$. $Q_1$ and $Q_2$ are given by its cyclic permutations. This means that three 
$Q_k|\Psi\rangle$ spans different Fock spaces. The reason is that, for $k=3$, once $D_6^{4}(1,2)$ is formed from $B_1$ and $B_2$, 
the simultaneous formation of $D_6^{4}(2,3)$ and/or $D_6^{4}(3,1)$ does not take place, because such a double (also triple) 
$D_6^{4}$ configuration cannot form a closed net of the oriented strings and junctions. The latent effect from the existence of $Q_3|\Psi\rangle$ 
on the three-baryon state $P|\Psi\rangle$ appears as the repulsive 3BI, $W(1,2;3)$. In the same way, 
combination of $D_6^{4}(2,3)$ and $B_1$ ($D_6^{4}(3,1)$ and $B_2$) leads to $Q_1|\Psi\rangle$ ($Q_2|\Psi\rangle$), which 
brings about $W(2,3;1)$ ($W(3,1;2)$) on $P|\Psi\rangle$. Thus the sum, $Q_T|\Psi\rangle=\sum_1^{3}Q_k|\Psi\rangle$, 
provides the SJM-3BI as the latent effect on $P|\Psi\rangle$, which is given by the cyclic sum of $W(i,j;k)$ as in Eq.~(4\cd2).  

The exotic tribaryon states can be attained also in one step from the ordinary $BBB$ system
 by creating two $B\bar{B}$ pairs as illustrated in the (right) of Fig.~7. Fusion of the strings of $q$ and $\bar{q}$ leads to 
the interjunctions, and the $BBB(B\bar{B})(B\bar{B})$ system turns into the exotic tribaryon $T_9^{7}$, shown in the (left) of Fig.~7, 
This state is the same with the one denoted by $A_9$ in ref. 11b).

\subsection{Effective two-body potential from three-baryon interaction in the baryonic matter} 
In order to evaluate the effect of the SJM-3BI in the baryonic matter, we employ a practical way, called the 
effective two-body potential (2BP) method, which enables us to calculate the effects from 3BI in a form of the additional contribution 
to the two-baryon potential. This method was developed by Hokkaido group in the study on the role of the 3BI due to 
the two-pion exchange via $\Delta$-excitation in nuclear matter,\cite{KAT74} and used in recent works on 3BI.\cite{GLMM89,ZLLM02} 
This is a way to obtain the effects of 3BI approximately, in place of full scale Bethe-Faddeev calculation with 3BI. 
The procedure in calculating the effective 2BP from the SJM-3BI are as follows. 
\begin{enumerate}
\item
First of all, effects of the 3BI appear under the influence of the primary interaction, the two-baryon interaction $V$, that is, 
its main effects are obtained with use of the wavefunction distorted by $V$. In baryonic matter, the distortion from the plane wave is 
remarkable at small inter-baryon distances, and is called the short-range correlation (abbreviated as SRC).
\item
Applying the method, we derive an effective 2BP, $U(1,2;\rho)$, from $W(1,2;3)$ acting in the baryonic matter, by taking the sum of the 
diagonal matrix elements of (i.e., by taking the average) over the occupied states of the 3rd baryon $B_3$. 
\item
By the procedure of taking this average, because the states of $B_3$ do not change, the effect from $W(1,2;3)$ results in 
the additional effect on the interaction between $B_1$ and $B_2$.    
\item
A remaining problem is to choose a functional form of $W(1,2;3)$ with use of three spatial variables 
(${\mbf r}_1,\;{\mbf r}_2$ and ${\mbf r}_3$), which are available here because of the universality (independence on flavor and spin) 
of the SJM-3BI.    
\end{enumerate}

\begin{eqnarray}
U(1,2;\rho)&\equiv& 
\int\; d{\mbf r}_3\sum_{\gamma}^{({\rm occ})}\langle \phi_{\gamma}(3)|W({\mbf r}_1,\;{\mbf r}_2\;;{\mbf r}_3)|
\phi_{\gamma}(3)\rangle f^{2}({\mbf r}_1-{\mbf r}_3)f^{2}({\mbf r}_3-{\mbf r}_2) , \nonumber \\
= &\rho& \int\; d{\mbf r}_3\;W({\mbf r}_1,\;{\mbf r}_2\;;{\mbf r}_3)f^{2}({\mbf r}_1-{\mbf r}_3)
f^{2}({\mbf r}_3-{\mbf r}_2) ,
\end{eqnarray}
where $f({\mbf r}_i-{\mbf r}_j)$ is a SRC function which includes the suppression of the relative wavefunction 
by the repulsive core for $B_i-B_j$ pair. Here we assume $W$ being local. $\phi_{\gamma}(3)$ is the plane wave state, 
whose spatial part is $\phi_{\gamma}(3)={\rm exp}(i{\mbf q}_{\gamma}\cdot{\mbf r}_3)/\sqrt{\Omega}$, $\Omega$ being the normalization volume. 
The sum over the states $\phi_{\gamma}$ simply gives the baryon density $\rho$,  
because of the universality of the SJM-3BI.   

The same procedure is applied to $W(2,3;1)$ and $W(3,1;2)$ leading to the effective 2BPs in the baryonic matter, 
$U(2,3;\rho)$ and $U(3,1;\rho)$, respectively.

For choice of the functional form of $W(1,2;3)$, at best, we can take into account its property that $W(1,2;3)$ is 
the interaction mediated by $B_3$ between $B_1$ and $B_2$ and is of short-range nature.  In this context 
we take the simplest form, the product of a function of $B_1-B_3$ distance and a function of $B_2-B_3$ distance, as follows. 
\begin{equation}
W({\mbf r}_1,{\mbf r}_2,;{\mbf r}_3)\equiv W_0\;g({\mbf r}_1-{\mbf r}_3)\;g({\mbf r}_2-{\mbf r}_3),
\end{equation}
where $g$ is a function such that $g(r_{ij}\rightarrow 0)\rightarrow 1$ and vanishes at $r_{ij}\gg \ell_s$ and $W_0\sim {\cal E}$. 
Here, for convenience of calculations, we choose a gaussian form for $g(r_{ij})$ with the range of $\eta_C=0.45-0.50$ fm in Eq.~(3.4). 
\begin{equation}
g({\mbf r}_i-{\mbf r}_j)={\rm exp}(-\lambda r_{ij}^{2}),\;\;\lambda\simeq \frac{1}{\eta_C^{2}}\simeq (4.9-4.0)\;{\rm fm}^{-2}.
\end{equation}

Denoting the Fourier transform of $g({\mbf r})$ as $G({\mbf q})$,
\begin{eqnarray}
g({\mbf r})&=&\frac{1}{(2\pi)^{3}}\int d{\mbf q}\;G({\mbf q})e^{i{\mbf q}\cdot{\mbf r}}, \\
G({\mbf q})&=&4\pi\int dr r^{2}e^{-\lambda r^{2}}j_0(qr)=(\pi/\lambda)^{3/2}e^{-q^{2}/4\lambda}.
\end{eqnarray}
After the transform of momentum variables, $U(1,2;\rho)=U({\mbf r}_{12};\rho)$ is expressed, as follows:  
\begin{equation}
U({\mbf r}_{12};\rho)=\frac{\rho W_0}{(2\pi)^{3}}\int d{\mbf q}\;e^{-i{\mbf q}\cdot({\mbf r}_1-{\mbf r}_2)}
   \int d{\mbf q}_1\; h({\mbf q}-{\mbf q}_1)G({\mbf q}_1)\;\int d{\mbf q}_2\; h({\mbf q}-{\mbf q}_2)G({\mbf q}_2).
\end{equation}
Here $h({\mbf q})$ is the Fourier transform of the SRC function squared $f^{2}$:
\begin{equation}
f^{2}({\mbf r})\equiv \int \;d{\mbf q}\;h({\mbf q})e^{-i{\mbf q}\cdot{\mbf r}}.
\end{equation}
It is to be noted that $U({\mbf r}_{12};\rho)$ is strongly density-dependent. 

Without the SRC, we have $h({\mbf q})=\delta ({\mbf q})$, and then the integral becomes  
\begin{eqnarray}
U_0({\mbf r}_{12};\rho)&=&\frac{\rho W_0}{(2\pi)^{3}}\int d{\mbf q}\;e^{-i{\mbf q}\cdot{\mbf r}_{12}}(G({\mbf q}))^{2} \\
&=&\rho W_0(\pi/2\lambda)^{3/2}{\rm exp}(-\lambda r^{2}/2).
\end{eqnarray}
Hereafter we write ${\mbf r}_{12}$ simply as ${\mbf r}$.\footnote{Making estimate of $U_0$, we have 
$U_0(r;\rho)=84 (\rho/\rho_0)e^{-2(r/{\rm fm})^2}$ MeV for $\lambda=4{\rm fm}^{-2}$ and $W_0=2$ GeV. This value is (5-3) times 
larger than that with SRC at $r\simeq (0-1)$ fm. So the SRC is important for calculating effects of three-body force.}

Now we calculate the effective 2BP ($U$). To take into account the SRC for $B_1$-$B_3$ and  for $B_2$-$B_3$, 
we calculate the two integrals with the same form in Eq.~(4\cd11). In doing so we need the SRC function $f$ extracted from 
the solution of the reaction matrix equation in nuclear matter.   

The SRC functions change state by state.  At low density of neutron matter, as the ${}^1 S_0$ wave contributes mainly, 
we can use the SRC function of this wave denoted as $f_S(r)$. As the density goes higher, this approximation is not allowed, and 
we should add both the contributions from ${}^1 S_0$ wave  and ${}^3 P_{J=0,1,2}$ wave. 
The latter depends on $J$ due to the effects from the noncentral potentials. For energy quantities, 
the contributions appear in the weighted average over ${}^3 P_{J=0,1,2}$, which is proportional to $(2J+1)$. 
We define the ${}^3 P$-wave SRC function:
\begin{equation}
f_P(r)\equiv \{ f(r;{}^3 P_{J=0})+3f(r;{}^3 P_{J=1})+5f(r;{}^3 P_{J=2})\}/9.
\end{equation}
As the ${}^1 S_0$-wave and the average ${}^3 P$ wave contribute with the statistical weight relevannt to the spin, 
we define the weigted averge of the SRC function as  
\begin{equation}
f_{{\rm wa}}(r)\equiv \{f_S (r)+3f_P (r)\}/4.
\end{equation}
These SRC functions are constructed using the wave functions obtained for OPEG-A potential\cite{OPEG} in the reaction matrix equation 
in the neutron matter.

In numerical calculations, we obtain the effective two-body potential from the SJM-3BI, using $f_S(r)$ or $f_{{\rm wa}}(r)$.
The damping of $f_S(r)$ at small distance is strong, but that of $f_P (r)$ is moderate. Also 
the damping of $f_{{\rm wa}}(r)$ is not so strong.

\begin{figure}[t]
 \parbox{\halftext}{
     \centerline{\includegraphics[width=65mm,height=9cm]{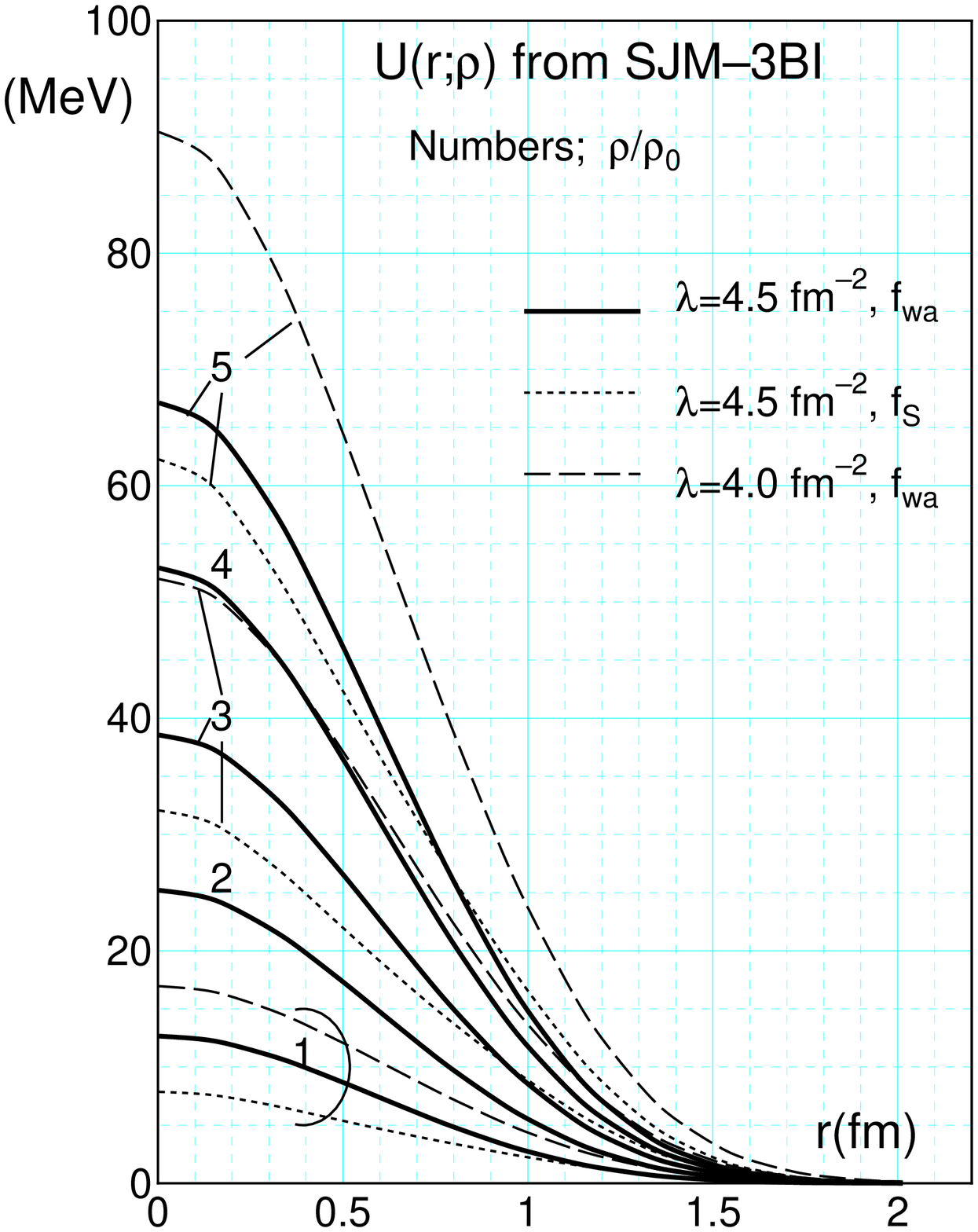}}
     \caption{Effective two-body potential $U(r;\rho)$ from universal repulsion of SJM-3BI.
               Numbers attached to the curves show $\rho/\rho_0$. Parameters of the force range $\lambda$ and the SRC function $f$ 
               are indicated in the figure.} 
 \label{fig:8}}
 \hspace{3mm}
 \parbox{\halftext}{
     \centerline{\includegraphics[width=65mm,height=8cm]{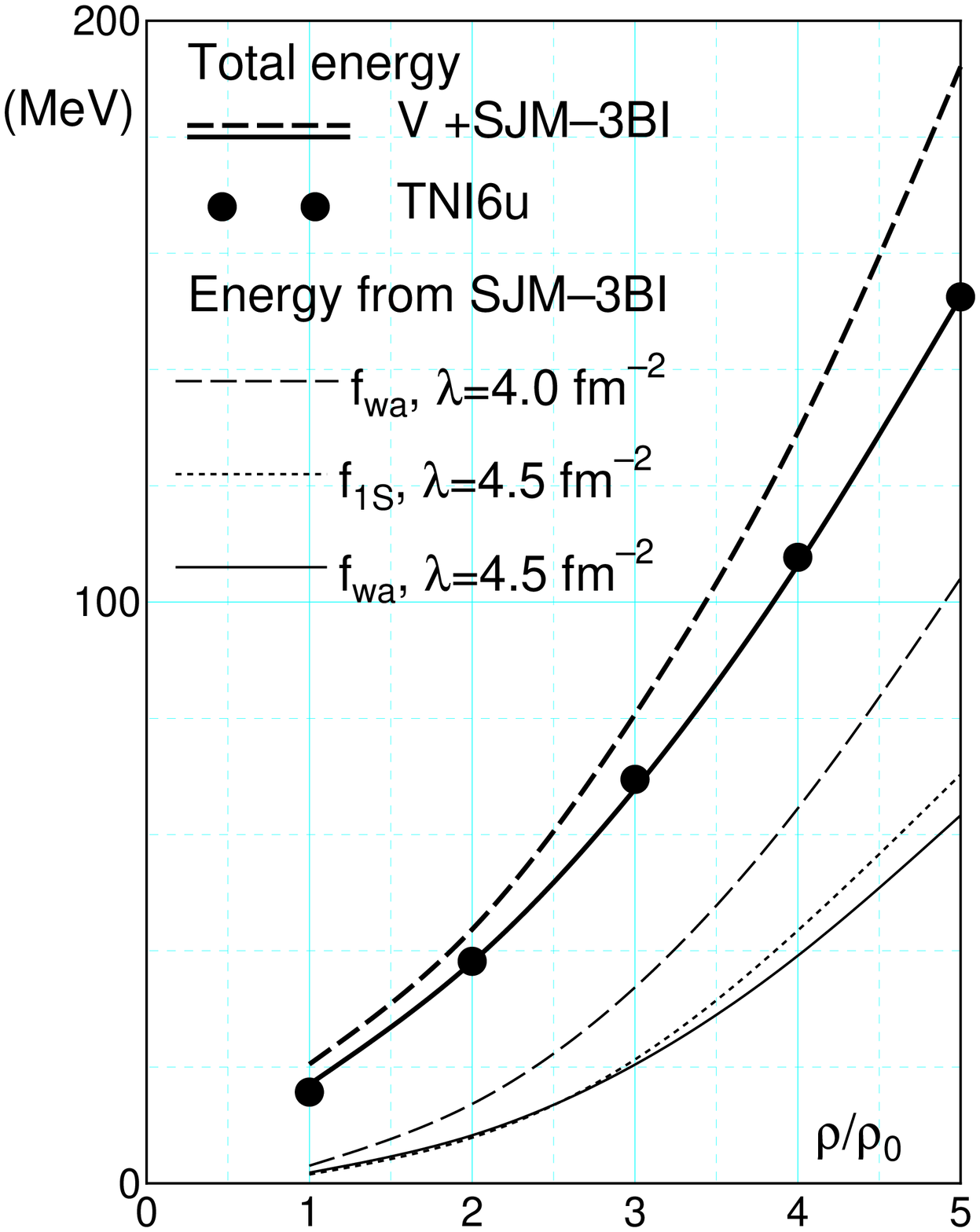}}
     \caption{Density dependence of effects from SJM-3BI on the energy/baryon in neutron matter.
               The lower three curves show contributions from SJM-3BI. The upper two curves show the total energy/baryon 
               in neutron matter including SJM-3BI, where two-body ($V$) contribution and 
               EOS of TNI6u (filled circles) are taken from ref. \citen{TNYT06}. } 
\label{fig:9}}
\end{figure}

We denote the Fourier transform $G$ modified by the SRC effect by $G_{\rm SRC}(q)$:
\begin{eqnarray}
G_{{\rm SRC}}(q)&\equiv& \int d{\mbf q}_1\; h({\mbf q}-{\mbf q}_1)G({\mbf q}_1)\\
&=&4\pi\int_0^{\infty} dr r^{2}f_L^{2}(r)e^{-\lambda r^{2}}j_0(qr),
\end{eqnarray}
where $h$ is substituted by its Fourier transform $h({\mbf q})=\int d{\mbf r}f_L^{2}(r){\rm exp}(i{\mbf q}\cdot{\mbf r})/(2\pi)^{3}$.
$G(q)_{{\rm SRC}}$ differs from $G(q)$ by including the SRC factor $f_L^{2}(r)$ in the integral.
Then the effective two-body potential is given as 
\begin{equation}
U(r;\rho)=\frac{\rho W_0}{2\pi^{2}}\int_0^{\infty}dq\;q^{2}j_0(qr)\{ G_{{\rm SRC}}(q) \}^{2}.
\end{equation}

In this paper the strength of the SJM-3BI is taken as $W_0=2$ GeV.  Results of numerical calculations for $U(r;\rho)$ 
are shown in Fig.~8 at the density $\rho/\rho_0=1-5$, where $\rho_0=0.17\;{\rm fm}^{-3}$ is the nuclear density. 
The effective two-body potential obtained from the SJM-3BI is a universal repulsion with the range of about 1 fm.
Although it is of the order of 10 MeV around the nuclear density $\rho_0$, it grows almost linearly as $\rho$ goes high. 
It is rather sensitive to the spread of the function $g(r)$; $U$ becomes stronger for a wider spread. 
The $U(r;\rho)$ obtained from the repulsive term of phenomenological 3BI used by Illinois group\cite{AP97} has 
a longer range and a weaker strength than those here shown.    

Adequacy of the 3BI obtained here should be examined in the energy properties of nuclei. We have studied  
the saturation curve of the symmetric nuclear matter, including this SJM-3BI in addition to a realistic two-nucleon 
potential with a repulsive core and the pionic 3BI of the Fujita-Miyazawa type, and obtained a reasonable saturation curve.\cite{TNTPSJreport}

\section{Effects on the equation of state of neutron matter and neuron star matter }

We calculate the effects from the SJM-3BI on the EOS (equation of state), energy/baryon, in neutron matter and then discuss the effects on 
the EOS in neutron star matter. As the SJM-3BI is universal, its first order effect in the perturbation (without the effects of the 
two-body correlation in Eq.~(4\cd3)) is simply given as
\begin{equation}
\frac{E_U^{(1)}}{N}=\frac{k_F^{3}}{6\pi}\int_0^{k_F} dk P(k)\{ 4\int dr r^{2}U(r;\rho)-2\int dr r^{2}U(r;\rho)j_0(2kr) \},
\end{equation}
where $N$ is the baryon number (here the neutron number), $k_F$ the Fermi momentum and the 1st (2nd) term in the curly bracket is 
the direct （exchange) contribution. The function $P(k)$ is the probability function of the relative momentum $k$, normalized 
as $\int_0^{k_F}dkP(k)=1$：
\begin{equation}
P(k)=\frac{24k^{2}}{k_F^{3}}\{1-(3/2)(k/k_F)+(1/2)(k/k_F)^{3} \}.
\end{equation}

The two-body correlation between $B_1$ and $B_2$ including the SRC effects are taken into account by multiplying $f_S(r)$ on the ${}^1 S_0$ wave 
and $f_P (r)$ on the ${}^3 P$ wave. Then the contribution to the energy/baryon from the SJM-3BI (the part of the $U$-term in Eq.~(4\cd3)) 
is given by
\begin{eqnarray}
\frac{E_U}{N} &=& \frac{k_F^{3}}{6\pi}\int_0^{k_F} dk P(k)\{ \int dr r^{2}U(r;\rho)(f_S(r)+3f_P(r)) \nonumber \\
              &+& \int dr r^{2}U(r;\rho)(f_S(r)-3f_P(r))j_0(2kr) \}.
\end{eqnarray}

Since the effective two-body potential $U$ has the density dependence being almost linear, the growth of $E_U/N$ is stronger than 
the linear density-dependence. The total energy /baryon is given by the sum of the contribution from the two-body potentilal $V$ 
and the contribution from the repulsive SJM-3BI ($E_U/N$), as shown in Fig.~9. For the former we use the result by Takatsuka et al.\cite{TNYT06}
The upper two curves show the total energy/baryon (EOS at zero temperature). The bold solid curve is for the case of the 
following parameters; 
the range of the function $g$, $\lambda=4.5\;{\rm fm}^{-2}$ (corresponding to the spread $\simeq 0.47$ fm), 
$W_0=2$ GeV and the $f_{\rm wa}$ for the SRC function. In this case we obtain almost the same EOS with TNI6u 
(shown by $\bullet$ in the figure)\cite{TNYT06} without adjustment of parameters. The predicted maximum mass of 
neutron stars is $M_{\rm max}\simeq 1.8M_{\odot}$, which is consistent with the observations. If we use a little wider range parameter, 
$\lambda=4.0\;{\rm fm}^{-2}$ (corresponding to the spread=0.5 fm), we can provide the stiffer EOS shown by the upper bold-dashed curve, 
which leads to $M_{\rm max}\simeq 2.0M_{\odot}$. A full accout concerning the EOS including the meson exchange 3BI  
will be  reported elsewhere.\cite{TNT07}    

Because this SJM-3BI is independent of flavor and spin, its repulsive effect is not influenced by the hyperon-mixing 
in the neutron star matter. Thus we can get rid of the dramatic softening of EOS, which otherwise leads to the contradiction 
with the observations. 
 
\section{Discussion}

Here we discuss on several related points.

\subsection{On the parameters in the SJM} 

In the arguments in the previous sections, we preferentially use the key number ${\cal E}\sim 2$ GeV, without direct use of the 
 parameters in the SJM.   Here we try to estimate the values of the parameters ($m_q$, $m_J$ and $m_{IJ}$) in the mass formula 
of the SJM, utilizing this number.  

Although $m_B$ (the masss of the single baryon $B$) is in considrable variation, its source is considered as mainly due to 
the difference in current masses of involved quarks. From the viewpoint of the SJM, since the binding mechanism is mainly determined 
by the strings and junctions, we have a ground for the statement that the baryon mass difference is mainly attributed to 
the current quark masses, if we disregard the effects of spin-dependent interactions such as the color magnetic interaction leading 
to the color hyperfine splitting. In the following we take the current quark mass$\simeq 0$, 
suitable for the $u$ and $d$ quarks. Thus we take $m_B\simeq 1$ GeV, which means $3m_q+m_J\simeq 1$ GeV in the SJM. 
First we prove the property, assumed in the previous sections, that $\delta\equiv 2m_q-m_{IJ}$ used in Eqs. (2\cd4) and (2\cd5) is minor 
compared with $m_B\sim 1$ GeV, which holds in a way irrelevant to the total length of the strings $L_{\rm min}$. Next we discuss 
its implications and make the estimate of respective parameters by taking two typical values of $L_{\rm min}$.

\begin{enumerate}
\item Substituting the relation $m_J=m_B-3m_q$ into the expression of ${\cal E}= 2m_J+3m_{IJ}\sim 2$ GeV $\simeq 2m_B$, 
we have $2m_q\sim m_{IJ}$. Then we obtain the relation $\delta=2m_q-m_{IJ}\sim 0$, namely, $\delta=O(0.1 $GeV).
This shows that $|\delta|\ll m_B$ holds, irrespective of $L_{\rm min}$.  
\item  The relation  $2m_q\sim m_{IJ}$ means that the energy for the fusion of $q$ and $\bar{q}$ accompanying strings is almost 
equal to the energy of the inter-junction string $m_{IJ}$. Owing to this property, we can show that all the states in the ${\cal M}$ series 
shown in Fig.~4 have the energy of about 2 GeV. Accordingly they can transform to each other without large energy transfer.  
\item By taking two typical values of $L_{\rm min}\simeq $1.0 fm and 1.5 fm, we give some estimate for the SJM parameters.
$L_{\rm min}\simeq $1.0 fm (1.5 fm) corresponds to the string length from the junction to the quark, $\ell_s\simeq L_{\rm min}/3 
\simeq$0.33 fm (0.5 fm) in the ground state of $B$. As noted already, the energy increase due to the string stretching is 
of the order of 1 GeV/fm. 
The string energy in $B$ is $3m_q \simeq \sigma L_{\rm min} \simeq$ 1 GeV (1.5 GeV), which leads to  $m_J\sim 0\;(-0.5$ GeV). 
The color Coulomb force gives the moderate attractive energy of about $-0.2$ GeV ($-0.1$ GeV).\cite{TSNM02} The effect is much weaker 
than those ($\sim -0.5$ GeV) obtained in calculations wiht use of the quark cluster model and the bag model,\cite{ITT} 
because the probability of close approach of the quarks is much smaller for the Y-shaped configuration. 
Energy contribution to $m_J$, being $\sim 0.2$ GeV ($-0.4$ GeV), is to be provided as energy concentration at the junction. 
Using the relation $2m_q\sim m_{IJ}$, we can estimate $m_{IJ}\sim$ 0.7 GeV (1 GeV), the energy needed for formation of an interjuction. 
\end{enumerate}
 
Thus there remains a problem how to understand the value of $m_J$ dependent on the choice of $L_{\rm min}$. As mentioned in \S 3, 
following the lattice QCD calculations,\cite{TSIMNO03c} the case of $L_{\rm min}\simeq 1.5$ fm seems favorable, since  
for $L_{\rm min}\simeq 1.0$ fm the string length $\ell_s$ is comparable with the width of the flux tube and 
the string-junction picture looks not so apparent. For the case of $L_{\rm min}\simeq 1.5$ fm, the negative energy of 
the order of $m_J\sim -0.4$ GeV is inevitable, since the string energy is considerably larger than $m_B$, but this feature 
is not unreasonable in the following sense. The negative $m_J$ means the concentration of negative energy at the junction 
in the ground state of $B$ without interjunction. In the exotic baryon states, however, the formation of new junctions inevitably 
accompanies the creation of interjunction(s) which needs the large positive energy in unit of $m_{IJ}\sim$ 1 GeV. 

Although the value of $m_J$ looks directly linked to the parameter $C_{3q}$ in Eq.~(2\cd2), estimate of $m_J$ from $C_{3q}$ is 
not possible because $C_{3q}$ contains an integration constant. Various effects to the energy of the ground state of $B$ 
not specified in the simple mass formula in the SJM would contribute to $m_J$, and this problem is beyond the scope of the present study. 
As noted already this problematic point is not obstacle in the present work, because  we have used the quantity ${\cal E}$ preferentially.   
 
\subsection{On the $BB$ repulsive core}

The $NN$ repulsive core is one of the main factors leading to the property that the nucleus with the low density is 
the lowest state in hadronic matter.\footnote{In the 1980's there was pointed out the possibility that the strange quark matter 
with almost equal numbers of $u$, $d$ and $s$ is the absolutely lowest state of hadronic matter. 
In a previous study with use of the quark cluster model and the bag model,\cite{ITT} we have shown that this possibility is not realized.} 
On the $BB$ repulsive core, there have been rarely studied by taking into account the confinement mechanism, which is 
the nonperturbative effect originating from the color degrees of freedom in QCD. In the studies in the quark cluster model, 
although the quark confinement is incorporated in a potentilal form between quarks, the interaction between clusters (baryons) 
is described in such a way that the confinement plays no substantial role as far as the color singlet is imposed to the total system, 
in spite of its significant contribution to the internal energy of the cluster of quarks (baryon). 
Since these studies have been developed to a stage that most of the existing 
data of the $BB$ system are well explained,\cite{OSYPTPS137,FSN07} the important findings are surely contained in the results 
of these works. Neverthless it seems strange that the effects of the confinement does not appear in any essential manner in the $BB$ interaction. 
We feel that something essential may be lacked in the recognition hihterto obtained. In the present paper we have shown 
a straightfoward way to understand the $BB$ repulsive core as originationg from the confinement. 

Recently the first realistic approach on nuclear force in lattice QCD calculations has been made in the quenched approximation.\cite{IAH06} 
Although the outside attraction and the inside repulsion of the $NN$ $S$-wave interaction have been reported, it is premature 
to discuss the results in quantative level, and we expect future progress. From the viewpoint of the present paper, it is desirable 
to see at what extent the process like $J\bar{J}$-pair creation takes part on the short-range interaction, although it will be a task beyond 
the quenched approximation. This problem is connected with the study to clarify whether exotic dibaryons exist or not.    
It seems quite natural that the lattice QCD simulation has not been performed yet concerning the exotic dibaryons,  
 since they appear through the virtual creation of the $B\bar{B}$ pair and that of the $q\bar{q}$ pairs, 
following the context of the present paper. 
To get observational evidence on these exotic dibaryon states, we expect future experiments.\footnote{Recently Imachi, Otsuki and Toyoda have studied 
unconventional hadrons including exotic ones, extending the string-junction model they developed in the 1970's.\cite{IOT07}. 
Further study by assigning the recently observed peak at the mass 4.43 GeV ($Z^{\pm}$)\cite{Bell07} as the exotic meson 
$M_4^{2}$ (e.g., $u\bar{d}c\bar{c}$ for $Z^{+}$), has suggested the appearance of the dibaryon $D_6^{4}$ here discussed at the mass 
about 4.3~GeV.\cite{Imachi07}} Concerning the dibaryon problems, it seems promising to obtain exclusive data with use of 
the anti-proton beam such as $\bar{p}+\hbox{}^{3} {\rm He}\rightarrow D_4^{6}+\gamma$ and $\bar{p}+\hbox{}^{4} {\rm He}\rightarrow D_4^{6}+N$.      

\subsection{On three-baryon interaction}                  

Hitherto in most cases, the three-body interaction in the $BBB$ system has been introduced as a supplementary effect to reproduce 
the data in a quantitative level. In recent years, its indispensable role in the nuclear system has been recognized progressively, 
concerning the saturation problem, the maximum mass of neutron stars, and the requirement to reproduce various kinds 
of precise measurements on few-nucleon systems. Nevertheless, it seems fair to ask why three-baryon interaction plays a 
decisive role in the fundamental aspects of the nucleus. We can accept that its importance is natural,  
if its short-range part originates from the confinement and its intermediate part comes from meson exchange, especailly pion exchange. 

\subsection{On exotic multi-baryon system}  
In this paper we have treated the exotic baryon systems of the baryon numbar $N=2,\;3$ in the SJM. What aspects of such exotic systems 
come about when $N$ becomes much larger? We can apply the same procedure as shown in Fig.~6 to the $N=4$ system, and find that its energy 
is higher than the ordinary value 4$m_B$ by $\sim 3{\cal E}$. Furthermore we can extend the treatment to the general case. 
Such an exotic multi-baryon system (abbreviated to EMBS), whose possibility was pointed out in Ref. 7c), 
have the energy higher than $N m_B$ of the ordinary baryon system by $(N-1){\cal E}$. Namely, EMBS has the saturation property of energy; 
${\cal E}$/baryon$\sim 2$ GeV/baryon. However to estimate the saturation density of EMBS is problematic, contrary to the unifom matter. 
The reason is that EMBS connected by strings, junctions and interjunction strings is flexible in spatial form like the ``chain". 
Therefore it seems difficult to consider EMBS beyond few-baryon systems, unless its spatial configuration is well defined.   

\section{Concluding remarks}

In this study we have shown a way to unifiedly understand the universal repulsive core of the $BB$ interaction and the universal 
repulsion of the three-baryon interaction, standing on the the string-junction model where the confinement mechanism in QCD is 
built in. 

The key points of recognition are summarized as follows. 
In order for the confinement mechanism to persisit at closest approach of baryons, the $B\bar{B}$-pair (more widely the 
 pair of the junction and anti-junction) excitation takes place and leads to the formation of a new closed string-junction net 
with exotic nature. For the $BB$ system the energy needed for the $B\bar{B}$-pair creation appears as the repulsive core with
 $\sim 2$ GeV height. For the $BBB$ system a half of the energy for the 2$B\bar{B}$-pair, being $\sim 2$ GeV, is regarded as 
the strength of the three-body potential. Participation of such exotic baryonic states brings about the latent effect 
on the ordinary baryon states, which appears as the short-range repulsion. Thus we can understand the universal 
(flavor- and spin- independent) short-range repulsion as originating from the confinement mechanism. 
Results of numerical calculations show that the universal repulsion of three-baryon interaction derived in the string-junction 
model provides us with the sufficiently stiff equation of state, which is consistent with the observations of 
neutron star masses, even when the hyperon-mixing occurs in the neutron star core. 

We have developed our studies on the assumption of the existence of exotic multi-baryon states. Hitherto experimental evidence  
about exotic hadrons has not been obtained yet.  Since there is no reason in QCD to deny the existence of hadrons other than the ordinary ones,  
this is a basic problem to be solved in future in hadron physics. Recently exotic hadrons have attracted much interest, 
exprimetally and theoretically,\cite{HADRON2005} and we hope that active studies now being in progress will elucidate such 
basic problems in future. 

Concerning the problem under investigation, the reasoning taken here will not lose validity even though 
exotic hadron states with narrow width are not observed.  The reason is that, e.g. for the $BB$ case, if there exist a group of states 
with a junction number differnt from that of the ordinary $BB$ case almost restricted to the small spatial space 
corresponding the core region of $BB$ and their energies are as high as 2 GeV, the ordinary $BB$ system can scarcely 
enter into such region. This implies the existence of the repulsive core.

\section*{Acknowledgements}
The author wishes to thank Dr.~Taktatsuka for providing motivation of this work and to thank Dr.~Otsuki for his interest in this work 
and thoughtful reply to author's questions on the string-junction model. Discussions on recent development in lattice QCD studies 
with Dr.~Suganuma have been very valuable.


\begin{thebibliography}{99}
\bibitem{SokenarXiv}R.~Tamagaki, Soryushiron Kenkyu (Kyoto), {\bf 115} (2007), 37.\\
                    R.~Tamagaki, arXiv nucl-th/0801.2289.
\bibitem{NYT02} S.~Nishizaki, Y.~Yamamoto and T.~Takatsuka, Prog.~Theor.~Phys. {\bf 108} (2002), 703.
\bibitem{Jastrow} R.~Jastrow, Phys. Rev. {\bf 81} (1951), 165.
\bibitem{TNTRCNP} T.~Takatsuka, S.~Nishizaki and R.~Tamagaki, Report in {\it RCNP Symposium on  
Few-Nucleon System and Baryon-Baryon Interaction}, held March 6-7, 2007, RCNP.
\bibitem{TSNM02} T.~T.~Takahashi, H.~Suganuma, Y.~Nemoto and H.~Matsufuru, Phys. Rev. {\bf D65} (2002), 114509.
\bibitem{Nambu73} Y. Nambu, {\it Proceedings of the Tokyo Symposium on High Enegy Physics}, July, 1973, p.221.
\bibitem{PTPS78} For reviews, for examples, \\
(a) Y.~Igarashi, M.~Imachi, T.~Matsuoka, K.~Ninomiya, S.~Otsuki, S.~Sawada and F.~Toyoda, Prog.~Theor.~Phys. Supplememt No.{\bf 63} (1978), 49. \\
(b) M.~Imachi, Domestic Journal of the Physical Society of Japan {\bf 33} (1978), No.~3, 207 (in Japanese).\\
(c) Chapter IV in ref. 7(a).
\bibitem{Tamagaki82} R.~Tamagaki, Bulletin of the Institute for Chemical Research, Kyoto Univ., Vol.{\bf 60}, No.2 (1982), 190.
\bibitem{TNYT06} T.~Takatsuka, S.~Nishizaki, Y.~Yamamoto and R.~Tamagaki, Prog. Theor. Phys. {\bf 115} (2006) 355.
\bibitem{Okubo78}For review, for example, S.Okubo, Prog.~Theor.~Phys. Supplememt No.{\bf 63} (1978), 1. 
\bibitem{IOT757677}  M.~Imachi, S.~Otsuki and F.~Toyoda, Prog.~Theor.~Phys.~(a)~{\bf 54} (1975), 280;~(b)~{\bf 55} (1976), 551;
~(c)~{\bf 57} (1977), 517.
\bibitem{IO77} M.~Imachi and S.~Otsuki, Prog.~Theor.~Phys. {\bf 58} (1977), 1660;~{\bf 58} (1977), 1657.
\bibitem{IO78} M.~Imachi and S.~Otsuki, Prog.~Theor.~Phys. {\bf 59} (1978), 1290.
\bibitem{TMNS01} T.~T.~Takahashi, H.~Matsufuru, Y.~Nemoto and H.~Suganuma, Phys. Rev. Lett.{\bf 86}(2001), 18.
\bibitem{Ichie03c} H.~Ichie, V.Bornyakov, T.~Streuer ang G.~Schierholtz, Nucl.~Phys.~{\bf A721} (2003),~899c.
\bibitem{TSIMNO03c} T.~T.~Takahashi, H.~Suganuma, H.~Ichie, H.~Matsufuru and Y.~Nemoto, Nucl.~Phys.~{\bf A721} (2003),~926c.
\bibitem{Miyazawa79} H.~Miyazawa, Phys. Rev. {\bf D20} (1979), 2953.
\bibitem{Nambu74} Y.~Nambu, Phys. Rev. {\bf D10} (1974), 4262.
\bibitem{SOTI05} H.~Suganuma, F.~Okiharu, T.~T.~Takahashi and H.~Ichie, Nucl. Phys. {\bf A755} (2005), 399c.
\bibitem{SIOT04} H.~Suganuma, H.~Ichie, F.~Okiharu and T.~Takahashi, {\it  Proceedings of the International \\ 
Workshop PENTAQUARK04}, held July, 2004 at Spring 8, ed. A.~Hosaka and T.~Hotta, (World Scientific, 2005), p. 414. 
\bibitem{TS04} T.~Takahashi and H.~Suganuma, Phys. Rev. {\bf D70} (2004), 074506.
\bibitem{OTW64} S.~Otsuki, R.~Tamagaki and M.~Wada, Prog.~Theor.~Phys. {\bf 32} (1964), 220.
\bibitem{OPEG} R.~Tamagaki, Prog.~Theor.~Phys. {\bf 39} (1968), 91; R.~Tamagaki and T.~Takatsuka, Prog. Theor. Phys. {\bf 105} (2001), 1059. 
\bibitem{OTY65} S.~Otsuki, R.~ Tamagaki and M.~Yasuno, Prog.~Theor.~Phys. Supplement, Extra Number (1965), 578.
\bibitem{Tamagaki67} R.~Tamagaki, Rev.~Mod.~Phys. {\bf39} (1967), 629.  
\bibitem{HT72} J.~Hiura and R.~Tamagaki, Prog.~Theor.~Phys. Supplememt No.{\bf 52} (1972), 25. 
\bibitem{Saito77} S.~Saito, Prog.~Theor.~Phys. Supplememt No.{\bf 62} (1977), 11. 
\bibitem{FujitaMiyazawa} J.~Fujita and H.~Miyazawa, Prog. Theor. Phys. {\bf 17} (1957), 360.
\bibitem{AP97} A.~Akmal and V.~R.~Pandharipande, Phys.~ReV. {\bf 56C} (1997), 2261.
\bibitem{HParnps00} H.~Heiselberg and V.~Pandharipande, Annu.~Rev.~Nucl.~Part.~Sci. {\bf 50} (2000), 481. 
\bibitem{KAT74} T.~Kasahara, Y.~Akaishi and H.~Tanaka, Prog. Theor. Phys. Supplment No.{\bf 56} (1974), 96.
\bibitem{GLMM89} P.~Prang\'e, A.~Lejeune, M.~Martzolff and J.-F.~Mathiot, \PR{C40,1989,1040}. 
\bibitem{ZLLM02} W.~Zuo, A.~Lejeune, U.~Lombardo and J.~F.~Mathiot, \NP{A706,2002,418}.
\bibitem{TNTPSJreport} T.~Takatsuka, S.~Nishizaki and R.~Tamagaki, Report 24aYE-7 in the 2007 autumn meeting of Phys. Soc. Japan. 
\bibitem{TNT07}  T.~Takatsuka, S.~Nishizaki and R.~Tamagaki, in preparation.
\bibitem{ITT} M.~Ishii and R.~Tamagaki, \PTP{87,1992,969}; \PTP{89,1993,657}; M.~Ishii, R.~Tamagaki and A.~Tohsaki, \PTP{92,1994,111}.
\bibitem{OSYPTPS137} Prog. Theor. Phys. Supplement No.{\bf 137} (2000), edited by M.~Oka, K.~Shimizu and K.~Yazaki.
\bibitem{FSN07} Y.~Fujiwara, Y.~Shimizu and C.~Nakamoto, {\it Progress in Particle and Nuclear Physics} {\bf 58} (2007), 439.
\bibitem{IAH06} N.~Ishii, S.~Aoki and T.~Hatsuda, arXiv nucl-th/0611096; Phys. Rev. Lett. {99,2007,022001}.
\bibitem{IOT07} M.~Imachi, S.~Otsuki and F.~Toyoda, Soryushiron Kenkyu (Kyoto), {\bf 115} (2007), 131.
\bibitem{Bell07} S.-K. Choi et al (The Belle Collaboration), hep-ex/07081790v2. 
\bibitem{Imachi07} M.~Imachi, private communication.
\bibitem{HADRON2005} Summary Talks given by E.~Klempt and by T.~Barns in {\it Proceedings of Hadron Spectroscopy, Eleventh Int. Conf.}, 
August, 2005, Rio de Jasneiro, eds. E.~Reis, C.~G\"obel, J.~S\'a Borges and J.~Magnin, AIP CP814 (2006).     
 

\end{thebibliography}
\end{document}